\documentclass[]{aa} 
 
\usepackage{graphicx, times, txfonts}

\newcommand{\msol}{\mbox{$M_{\odot}$}} 
 
\newcommand{\msolyr}{{$M_{\odot}$}\,yr$^{-1}$} 
\newcommand{\mdot}{$\dot{M}$}
\newcommand{\lsol}{\mbox{$L_{\odot}$}} 
 
\newcommand{\kks}{K km s$^{-1}$} 
\newcommand{\ks}{km s$^{-1}$}

\begin{document}

\title{
First detection of rotational CO line emission in an RGB star
\thanks{
Based on observations made with ESO telescopes at the La Silla Paranal
Observatory under programme ID 091.D-0073 (ESO time) and 091.F-9322 (Swedish time).
Based on observations with the Atacama Pathfinder EXperiment (APEX) telescope. 
APEX is a collaboration between the Max Planck Institute for Radio Astronomy, 
the European Southern Observatory, and the Onsala Space Observatory.
Based on observations carried out with the IRAM 30m Telescope under programme 183-11. 
IRAM is supported by INSU/CNRS (France), MPG (Germany) and IGN (Spain).
}
}  
 
\author{ 
M.~A.~T. Groenewegen 
}

\institute{ 
Koninklijke Sterrenwacht van Belgi\"e, Ringlaan 3, B--1180 Brussel, Belgium \\ \email{martin.groenewegen@oma.be}
} 
 
\date{received: 2013,  accepted: 2013} 
 
\offprints{Martin Groenewegen} 
 
 
\abstract {
For stars with initial masses below $\sim$ 1 \msol, the mass loss during the first red giant branch (RGB) phase 
dominates mass loss in the later asymptotic giant branch (AGB) phase. 
Nevertheless, mass loss on the RGB is still often parameterised by a simple Reimers law in stellar evolution models.
}
{To try to detect CO thermal emission in a small sample of nearby RGB stars with reliable \it Hipparcos \rm parallaxes 
that were shown to have infrared excess in an earlier paper.
}
{A sample of five stars was observed in the CO J=2-1 and J=3-2 lines with the IRAM and APEX telescopes.
}
{One star, the one with the largest mass-loss rate based on the previous analysis of the spectral energy distribution, was detected.
The expansion velocity is unexpectedly large at 12 \ks.
The line profile and intensity are compared to the predictions from a molecular line emission code.
The standard model predicts a double-peaked profile, while the observations indicate a flatter profile.
A model that does fit the data has a much smaller CO envelope (by a factor of 3), and a CO abundance that is two times larger and/or 
a larger mass-loss rate than the standard model.
This could indicate that the phase of large mass loss has only recently started. 
}
{The detection of CO in an RGB star with a luminosity of only $\sim$1300 \lsol\ and a mass-loss rate as low as a few 10$^{-9}$ \msolyr\ is 
important and the results also raise new questions.
However, ALMA observations are required in order to study the mass-loss process of RGB stars in more detail, both for reasons of sensitivity 
(6 hours of integration in superior weather at IRAM were needed to get a 4$\sigma$ detection in the object with the largest detection probability), 
and spatial resolution (to determine the size of the CO envelope).
}


\maketitle

\section{Introduction} 

Stars with initial masses of $\la$ 2.2 \msol\ will go through an evolutionary phase called the first red giant branch (RGB), 
where stars reach high luminosities ($\log (L/ L_\odot ) \sim 3$).
For the lowest initial masses ($\la$ 1 \msol), the total mass lost during 
the RGB phase dominates that of the AGB (asymptotic giant branch) phase, and
therefore it is equally important to understand how this process develops. 
In stellar evolutionary models, the RGB mass loss is often
parameterised by the Reimers law (1975) with a scaling parameter.

In Groenewegen (2012; hereafter G12) a sample of 54 nearby RGB stars
was studied. In particular their spectral energy distributions (SEDs)
were constructed and fitted with a dust radiative transfer model in order to determine the mass-loss rate.
This is complementary to most of the studies on the subject that are based on RGB stars in clusters (see extensive references in G12).
Among the 54 stars, 23 stars are found to have a significant infrared excess, 
which is interpreted as mass loss. In the range 265 $< L < 1500 \;\lsol$, 
22 stars out of 48 experience mass loss, which supports the notion of episodic mass loss.

In the derivation of the (dust) mass-loss rate an expansion velocity ($v_{\infty}$) of 10~\ks\ was adopted.
The two main reasons for attempting to detect thermal CO emission is to determine 
the expansion velocity, and to get an independent estimate of the mass-loss rate.
In Sect.~2, the observations are described.
In Sect.~3, the results are shown, and interpreted using the model introduced in Sect.~4.
The results are discussed in Sect.~5.

\section{The observations} 

The CO(2-1) observations were obtained on 18-21 February 2012 with the IRAM telescope, located in the Sierra Nevada near Grenada, Spain.
Both polarisations of the EMIR E230 receiver (Carter et al. 2012) were tuned to the CO J=2-1 line 
and both the VESPA and FTS backends were connected. The results are consistent, and only the results obtained from VESPA are discussed.
At that frequency the beam is 10.7\arcsec\ (FWHM), and a main-beam efficiency of $\eta_{\rm mb}$= 0.58 is adopted.
%
The weather was at times extremely good (pwv $<$0.5 mm), and average
system temperatures of $T_{\rm sys} \sim$ 240~K ($T_{\rm mb}$ scale) and zenith optical depths $\tau \sim$ 0.05 were obtained for some stars.
Wobbler switching was used with a throw of 120\arcsec.


The CO(3-2) observations were obtained in service mode on 11 dates between April 14 and July 11, 2013, with the APEX telescope  
located in the Atacama dessert in Chile.
The XFFTS (eXtended bandwidth Fast Fourier Transform Spectrometer) backend (see Klein et al. 2012) was connected to the
APEX-2 receiver (Risacher et al. 2006)\footnote{see http://gard04.rss.chalmers.se/APEX\_Web/index.htm} and tuned to the CO J=3-2 line.
At that frequency the beam is 17.3\arcsec\ (FWHM), and a main-beam efficiency of $\eta_{\rm mb}$= 0.73 is adopted. 
The weather was average (pwv 1.0-1.5 mm). Wobbler switching was used with a throw of 50\arcsec.

Five stars were observed from G12, selected because of their visibility and having the highest mass-loss rates.
The sources HIP 44126, 53449, 67665, and 88122 were observed with IRAM, and HIP 44126, 53449, and 61658 with APEX.
Information on the stars is collected in Table~\ref{Tab-sample}. All entries are taken from G12, except the 
heliocentric radial velocities listed by SIMBAD, that come from Famaey et al. (2009, HIP 88122) and Famaey et al. (2005; the other four stars).
The mass-loss rates quoted are derived in G12 based on the modelling of the spectral energy distribution with a dust radiative transfer code 
and assuming an expansion velocity 
of 10 \ks\ (a typical expansion velocity, at least for AGB stars) and a dust-to-gas ratio of 1/200.

\section{The results}
\label{theresults}

\begin{figure} 

\resizebox{\hsize}{!}{\includegraphics[angle=0]{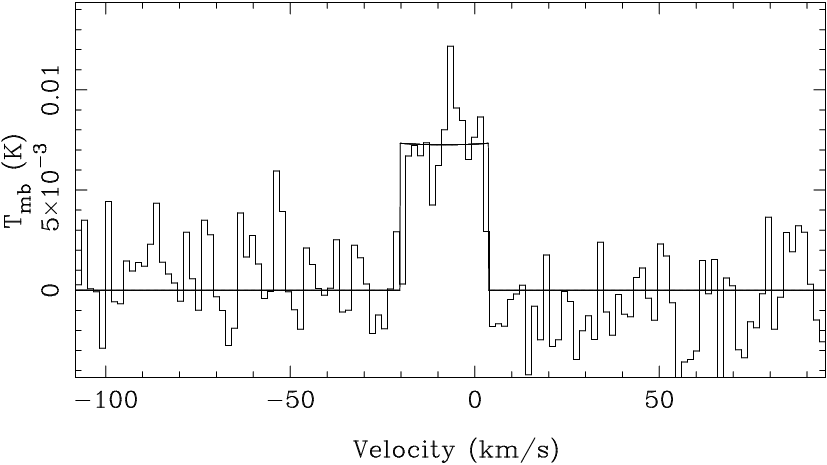}}
\resizebox{\hsize}{!}{\includegraphics[angle=0]{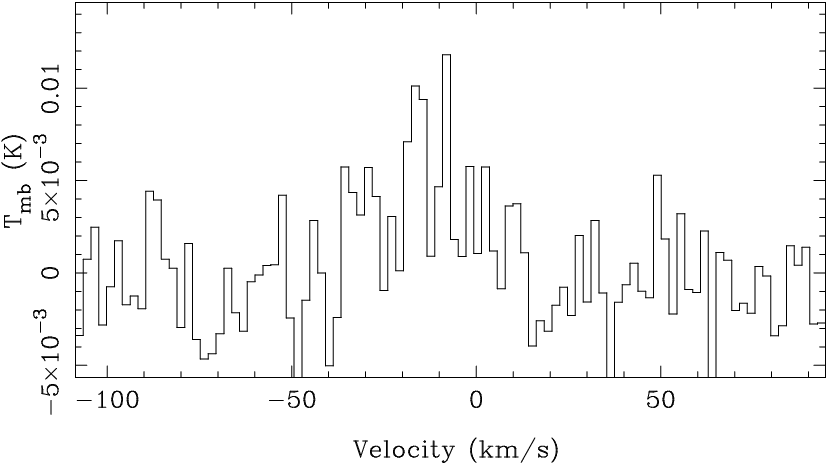}}
\resizebox{\hsize}{!}{\includegraphics[angle=0]{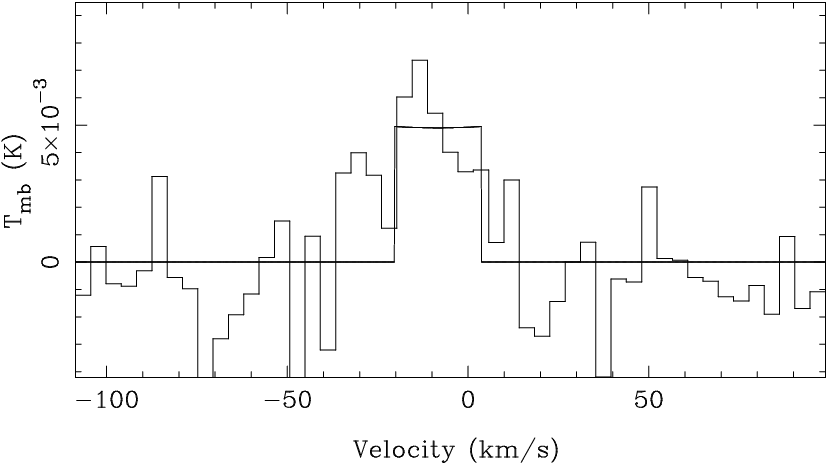}}
\caption[]{ 
HIP 53449. 
Top: observed CO(2-1) line profile  at a velocity resolution of about 2 \ks\ (histogram), and a typical fit of a shell profile (line, see text).
Middle: observed CO(3-2) line profile at a velocity resolution of about 2 \ks.
Bottom: observed CO(3-2) line profile at a lower resolution of about 4 \ks\ (histogram), and a typical fit of a shell profile (line, see text).
The velocity scale is heliocentric (the conversion from LSR velocity is done within CLASS).
} 
\label{fig-hip53449} 
\end{figure}

One star is detected, in both transitions. Figure~\ref{fig-hip53449} shows the spectra for HIP 53449, for a channel separation of about 2 \ks.
Detailed information on the observation results for all five objects are listed in Table~\ref{Tab-results}.

The J=2-1 line profile is 
well determined, and can best be described by a flat-topped profile.
There appears to be a narrow peak almost exactly at the stellar velocity of $-8.3$ \ks, but this probably fortuitous. 
In the FTS spectrum the peak is also present but even less clear, and in the original data at the highest velocity resolution 
(the FTS with a channel separation of 0.25 \ks, rms noise of 6.5 mK) the peak is not evident. 
The J=3-2 profile is much noisier and does not provide much insight. In the standard model discussed below a double-peaked profile 
is expected for both transitions.

\begin{table*}


\caption{The RGB sample.}

  \begin{tabular}{rrrrrrrrrrrr}
  \hline \hline
HIP   &   parallax     & $V_{\rm helio}$ &  $M_{\rm V}$ &  $A_{\rm V}$ & $V$ & $(V-I)$ & spectral &  $T_{\rm eff}$ &  $L$    & \mdot    & \mdot/$D^2$ \\
      &   (mas)        &  (\ks)        &             &             &     &         & type     &    (K)        & (\lsol) & (\msolyr) &  (\msolyr/ kpc$^{-2}$) \\ 
\hline

44126 & 5.33 $\pm$ 0.44 & $ 22.99 \pm 0.23$ & $-$0.16 & 0.09 & 6.30 & 2.48 & M4III   &  3600 &  528  $\pm$ 16  &  2.54 $\cdot$ 10$^{-9}$ &  72 $\cdot 10^{-9}$ \\
53449 & 8.42 $\pm$ 0.37 & $ -8.29 \pm 0.25$ & $+$0.49 & 0.05 & 5.91 & 3.50 & M5.5III &  3300 & 1346  $\pm$ 44  &  4.41 $\cdot$ 10$^{-9}$ & 313 $\cdot 10^{-9}$ \\
61658 & 6.64 $\pm$ 0.31 & $-15.60 \pm 0.36$ & $-$0.29 & 0.08 & 5.68 & 2.32 & M3III   &  3600 &  601  $\pm$ 22  &  1.05 $\cdot$ 10$^{-9}$ &  46 $\cdot 10^{-9}$ \\
67665 & 5.43 $\pm$ 0.20 & $-44.21 \pm 0.25$ & $-$1.65 & 0.08 & 4.76 & 1.63 & K5III   &  3600 & 1895  $\pm$ 73  &  3.81 $\cdot$ 10$^{-9}$ & 112 $\cdot 10^{-9}$ \\
88122 & 5.71 $\pm$ 0.25 & $ -9.47 \pm 0.80$ & $-$0.63 & 0.10 & 5.69 & 1.69 & M0III   &  4000 &  351  $\pm$ 15  &  1.77 $\cdot$ 10$^{-9}$ &  58 $\cdot 10^{-9}$ \\

\hline
\end{tabular}

\label{Tab-sample}
\end{table*}

\begin{table}
\setlength{\tabcolsep}{1.2mm}


\caption{The observed results.}

  \begin{tabular}{cccccccccc}
  \hline \hline
HIP   & $t_{\rm int}$ & $T_{\rm sys}$ & $\tau$ &  rms  & $\Delta v$ & $T_{\rm peak}$ & $\int T dv$ \\
      &  (min.)   & (K)         &        &  (mK) & (\ks)      &    (mK)       &   (\kks)    \\
\hline
\multicolumn{8}{c}{ IRAM }  \\

44126 & 277 & 258 & 0.094 & 3.28 & 1.63 & - & - \\ 
53449 & 430 & 240 & 0.047 & 2.43 & 1.63 & 8.0 $\pm$ 0.5  & 0.18 $\pm$ 0.01  \\ 
67665 & 392 & 234 & 0.036 & 2.61 & 1.63 & - & - \\ 
88122 & 188 & 278 & 0.097 & 4.10 & 1.63 & - & - \\ 

\multicolumn{8}{c}{ APEX }  \\

44126 & 159 & 452 & 0.252 & 5.32  & 2.12 & - & - \\ %
53449 & 525 & 430 & 0.156 & 2.84  & 2.12 & - & - \\
      &     &           &       & 2.01  & 4.42 & 5.3 $\pm$ 0.8 & 0.12 $\pm$ 0.02  \\ %
61658 & 349 & 434 & 0.212 & 3.70  & 2.12 & - & - \\ %

\hline
\end{tabular}
\tablefoot{Hipparcos numbers, integration times (in minutes), system temperatures, zenith optical depths, rms noise at the listed channel separation, 
and peak temperatures and integrated intensities of the line profiles. All temperatures are on the main-beam scale.  }
\label{Tab-results}
\end{table}

A shell-type profile, the one that is used in the CLASS software package\footnote{see http://www.iram.fr/IRAMFR/GILDAS}, is fitted to the data:
\begin{equation}
T({\rm v}) = \frac{A}{\Delta V} \;\;\; \frac{1 + 4 \; H \; (({\rm v - v_0})/\Delta V)^2}{1 + H/3},
\end{equation}
with v$_0$ and $\Delta V$ the central velocity, and the full width at zero intensity (both in \ks); 
$A$ is the area under the profile (in \kks); and $H$ represents the horn-to-centre ratio 
($-1$ for parabolic optically thick lines, $0$ for flat-topped profiles, and a large positive number for double-peaked lines).

The data is still too noisy to allow all parameters to be fitted. For v$_0$ fixed to $-$8.3 \ks, and for various fixed line profiles $\Delta V$ 
is found (and then fixed) to be 24 $\pm$ 0.7 \ks\ based on the J=2-1 line. The best-fit J=2-1 line profile is near $-$0.4, so towards a 
parabolic shape,  but the area under the curve and to a lesser extent the peak temperature listed in Table~\ref{Tab-results} are largely 
independent of the line shape. Figure~\ref{fig-hip53449} shows the fit for a flat-topped profile in the top and bottom panels with 
the numerical values listed in Table~\ref{Tab-results}. The error bars are derived from a Monte Carlo simulation where the shell-type 
profile was fitted to 1001 profiles generated from the observed one, but adding the rms noise.
It is remarked that $T_{\rm peak}$ is not fitted independently but derived from $T_{\rm peak} =  \frac{A}{\Delta V \; (1 + H/3) } $.

\begin{figure} 
\resizebox{\hsize}{!}{\includegraphics[angle=0]{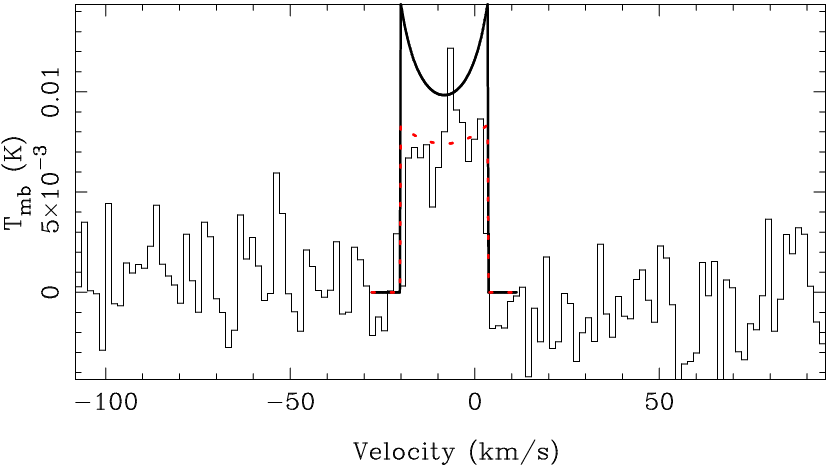}}
\resizebox{\hsize}{!}{\includegraphics[angle=0]{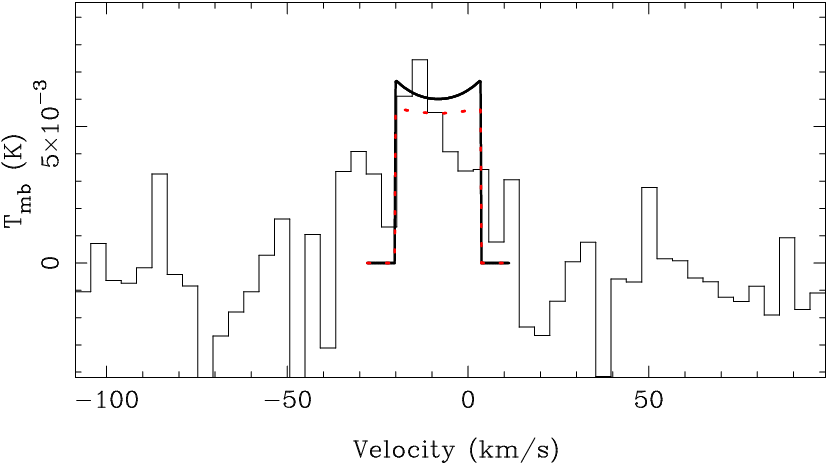}}
\caption[]{ 
Observed line profiles (histograms) for the J=2-1 (top) and J=3-2 (bottom) transitions.
The velocity scale is heliocentric.
The standard model is shown as the black solid line, the non-standard model as the red dotted line (see text for details). 
} 
\label{fig-hip53449-std} 
\end{figure}

\section{A model} 
\label{S-model}

To help in the interpretation of the observations, the molecular line emission code of Groenewegen (1994a,b) was used. 
This calculates the line profiles of thermal emission of molecules in an expanding spherically symmetric envelope.

The expansion velocity is set to 12~\ks. Other parameters (luminosity, stellar radius, effective temperature, grain properties) are taken from G12.
In the standard model the mass-loss rate is $5.3 \cdot 10^{-9}$ \msolyr, i.e. the value in Table~\ref{Tab-sample} for an 
expansion velocity of 12~\ks\ instead of 10~\ks\ and the dust-to-gas ratio is the one assumed in deriving 
this mass-loss rate from the dust modelling, namely 0.005.
The CO abundance relative to H$_2$ is set to $f_{\rm CO} = 2 \cdot 10^{-4}$ (Ramstedt et al. 2008).
It was verified that cooling by water can be neglected. Following the argumentation in Groenewegen (1994a, near Eq.~16) water cooling can only be important 
up to a distance of about $0.19 \cdot 10^{15}$ cm for this mass-loss rate and expansion velocity (it will be dissociated to OH farther out), 
which is about 2 stellar radii, too close to influence the intensity of the CO lower transitions.
A velocity law of the form $v/v_{\infty} = \left( 1 - R_{0}/ r \right)^{\gamma}$ is adopted in Groenewegen (1994a,b), 
with $R_{0} = 2 R_{*}$ and $\gamma = 0.5$ in the standard model.

The relative abundance of CO in the envelope is described by the formula proposed by Mamon et al. (1988),
\begin{equation}
X_{\rm CO} = e^{-\ln (2) \; (r/r_{1/2})^{\alpha}},
\end{equation}
where $r_{1/2}$ and $\alpha$ depend on $\dot{\rm M}$ and $v_{\infty}$ and are
tabulated for $f_{\rm CO} = 4 \cdot 10^{-4}$ by Mamon et al. (1988), and a fit formula is presented in Stanek et al. (1995).
This formulation gives $r_{1/2} = 7 \cdot 10^{15}$ cm and $\alpha$ = 1.5.
The outer radius of the envelope is set to where the relative CO abundance reaches 1\%, or about $2.4 \cdot 10^{16}$ cm.

Figure~\ref{fig-hip53449-std} compares the observation to the standard model (the black solid line). 
The peak intensity and integrated intensity are in remarkable agreement 
(about 30\% for the 2-1 line, and better for the 3-2 line).
The obvious discrepancy is in the line shape, which is predicted to be double-peaked, while the observations show otherwise. 
At the known Hipparcos distance of 119 pc, the value of the outer radius (about 3.5 times $r_{1/2}$) corresponds to an angular diameter of 27\arcsec, 
compared to the FWHM beam sizes of 11 and 17\arcsec.
The CO shell of the model is resolved by the beams, while the observations suggest otherwise.
One way to resolve this discrepancy would be to reduce $r_{1/2},$ normally implying lower mass-loss rates. The mass-loss rate 
of a few  $10^{-9}$ \msolyr\ is already smaller 
than the lowest value listed by Mamon et al. ($1 \cdot 10^{-8}$ \msolyr), and in fact for smaller and smaller mass-loss 
rates a constant value for $r_{1/2}$ is reached asymptotically which is the value already adopted in the standard model.

The red dotted line in Fig.~\ref{fig-hip53449-std} shows a good fit to the data, reached after some test runs. 
The parameter $r_{1/2}$ is reduced to  $3 \cdot 10^{15}$ cm 
(outer radius  $1 \cdot 10^{16}$ cm)  to give only a slight double-peaked profile, in better agreement with the data. Making the envelope smaller 
will result in lower intensities,  and this is compensated for by doubling the CO abundance to $f_{\rm CO} = 4 \cdot 10^{-4}$, 
which is still a reasonable value. 
The same effect could be reached by doubling the mass-loss rate and halving the dust-to-gas ratio.
Reducing the envelope size even more would result in a flat-topped and eventually parabolic profile, but would require increasing the CO abundance 
and/or mass-loss rate as well in order to match the observed intensity.

The shape of the velocity law has little influence on this result. Increasing the velocity gradient by essentially a factor of two far from the star 
(i.e. changing $\gamma$ from 0.5 to 1) results in nearly identical line shapes and intensities at the line centre that are lower by 7-10\%.
Similarly, arbitrarily reducing the kinetic temperature by 20\% has very little effect on line shapes and intensities. 

\section{Discussion}

From a sample of 54 nearby RGB stars, 23 of which were shown to have an infrared excess in G12, one star of five observed is detected in CO.
It is not easy to distinguish between an RGB and an early-AGB star. Specifically for HIP 53449, the private version of the 
PARAM tool\footnote{http://stev.oapd.inaf.it/cgi-bin/param} (da Silva et al. 2006), based on the star's effective temperature, 
distance, and luminosity, suggests that the star is five times more likely to be on the RGB than the E-AGB (Girardi, private communication)

Even {\em if} the star were an E-AGB star and not an RGB star its properties would still be exceptional.
Olofsson et al. (2002) present a complitation of the properties of 65 irregular and semiregular oxygen-rich variables based on CO observations. 
The mass-loss rates range from 2 to 80 $\cdot 10^{-8}$ \msolyr, with a median value of $20\cdot 10^{-8}$ \msolyr.
The object with the lowest mass-loss rate listed is L$^2$ Pup ($2 \cdot 10^{-8}$ \msolyr, at 85 pc for an assumed luminosity of 4000 \lsol). 
Taking its Hipparcos parallax (corresponding to 64 pc) one obtains an improved estimate of $\sim 1 \cdot 10^{-8}$ \msolyr\ and a 
luminosity of 2300 \lsol.
This is still considerably more than the $\sim0.5 \cdot 10^{-8}$ \msolyr\ at 1300 \lsol\ that is derived here for HIP 53449.
The next largest mass-loss rate in the Olofsson et al. sample is BI Car with $3 \cdot 10^{-8}$ \msolyr, but no parallax is known.
The next object is $\theta$ Aps, where the parallax is in good agreement with the assumed distance, so this object has a mass-loss rate of 
$4 \cdot 10^{-8}$ \msolyr\ at 4000 \lsol.
The mass-loss rate in HIP 53449 is extremely low, and at luminosities lower than previously detected.

From simple arguments one expects the peak line intensity to scale with $T_{\rm peak} \sim $ \mdot/$D^2$ (Knapp \& Morris 1985).
The one star detected has in fact the largest value of this parameter (see the last column of Table~\ref{Tab-sample}).
The non-detections of the other four objects are then expected given the lower value of \mdot/$D^2$ and the rms noise in the spectra.
This simple comparison assumes that all stars have the same expansion velocity and dust-to-gas ratio, which may not be true.

Nevertheless, the single detection of the star with the largest detection probability 
(under exceptional weather conditions with a 30m single-dish telescope) 
shows that in order to detect mass-loss rates in more objects requires significantly 
lower rms noise levels, which can only be achieved with ALMA.

Another question that can only be addressed by observing a larger sample is the width of the profile.
If it is interpreted as the outflow velocity of an expanding wind, then this velocity (of 12~\ks) is
larger than expected. This value is typical of an average AGB star, while HIP 53449 has a luminosity of only 1300~\lsol.

The high spatial resolution of ALMA observations could address one
of the other interesting findings, namely that the observed line shapes suggest
a smaller CO envelope than predicted. New CO photo-dissociation calculations for mass-loss rates 
lower than considered by Mamon et al. (1988) would be of interest to see if smaller values for $r_{1/2}$ are found.
If the CO envelope is indeed smaller then this could imply that the phase of large mass loss only started recently.

Groenewegen (2012) found that in the range 265 $< L < 1500 \;\lsol$ only 22 stars out of 48 show mass loss,
which supports the notion of episodic mass loss proposed by Origlia et al. (2007).
As a test, a model was constructed where the CO abundance was constant with radius, and the outer radius was decreased until the 
profiles matched the observations.
For $f_{\rm CO} = 2 \cdot 10^{-4}$ this is $5 \cdot 10^{15}$ cm and for $f_{\rm CO} = 4 \cdot 10^{-4}$ this is $3 \cdot 10^{15}$ cm (or 1.7\arcsec).
There is a feedback on the SED modelling because this was originally performed assuming a large outer radius 
(several 1000 stellar radii) while the CO shell in this test model is only 30-50 stellar radii. Redoing the SED fitting with a 
smaller outer radius will increase the mass-loss rate estimate to $8.1 \cdot 10^{-9}$  \msolyr, which in turn leads to an even smaller 
CO shell of $2.5 \cdot 10^{15}$ cm (for $f_{\rm CO} = 4 \cdot 10^{-4}$).
For a velocity of 12~\ks\ a radial distance of $(2.5-5) \cdot 10^{15}$ cm corresponds to a timescale of 65-130 years.
This is a very short timescale, much shorter than the lifetime on the RGB. 
It would imply that HIP 53449 is caught in the moment where it has just started its phase of large mass loss, which is statistically unlikely.

The observations and the single detection presented here show that detailed studies of the mass-loss process around 
nearby RGB stars are possible, but require the use of ALMA to increase the number of stars detected and to probe the geometry of the envelope.

\acknowledgements{  
I would like to thank Leo Girardi (Padova Observatory) for computing the RGB vs. E-AGB probability.
This research has made use of the SIMBAD database, operated at CDS, Strasbourg, France. 
}

{}

\end{document}